\documentclass[11pt,a4paper]{article}
\usepackage[T1]{fontenc}
\usepackage[utf8]{inputenc}
\usepackage{lmodern,microtype}
\usepackage[a4paper,margin=25mm]{geometry}
\usepackage{setspace}
\usepackage{amsmath,amssymb,amsthm,mathtools}
\usepackage{booktabs,tabularx,array,enumitem,needspace,etoolbox}
\usepackage{tikz,xurl,xcolor}
\usepackage{flafter,placeins}
\usetikzlibrary{arrows.meta,positioning,calc}
\usepackage[numbers,sort&compress]{natbib}
\usepackage[hidelinks,breaklinks=true]{hyperref}
\usepackage[capitalise,noabbrev]{cleveref}
\newif\iflinenumbers
\usepackage{lineno}
\setlist[enumerate]{label=(\arabic*),leftmargin=2em,itemsep=3pt}
\setlist[itemize]{leftmargin=1.5em,itemsep=2pt}
\newtheorem{theorem}{Theorem}[section]
\newtheorem{introresult}{Theorem}

\newtheorem{lemma}[theorem]{Lemma}
\newtheorem{proposition}[theorem]{Proposition}
\newtheorem{corollary}[theorem]{Corollary}
\theoremstyle{definition}
\newtheorem{definition}[theorem]{Definition}
\theoremstyle{remark}

\newcommand{\statementstart}[1]{\BeforeBeginEnvironment{#1}{\Needspace{4\baselineskip}}}
\forcsvlist{\statementstart}{introresult,theorem,lemma,proposition,corollary,definition}
\newcommand{\N}{\mathbb N_0}

\newcommand{\Prob}{\mathbb P}

\newcommand{\one}{\mathbf1}

\newcommand{\floor}[1]{\left\lfloor #1\right\rfloor}

\newcommand{\HALT}{\mathsf{HALT}}
\newcommand{\PostU}{\mathcal U}

\newcommand{\Phalt}{P_{\mathrm{halt}}}
\newcommand{\nchg}{m_{\mathrm{chg}}}

\AtBeginEnvironment{table}{%
	\setlength{\abovecaptionskip}{0pt}%
	\setlength{\belowcaptionskip}{5pt}%
	}
\hypersetup{pdftitle={What Can Postselection on Halting Decide?},pdfauthor={Saman Farhat and Farzad Didehvar}}
\begin{document}
	\nocite{*}
	\begin{center}
		{\LARGE\bfseries What Can Postselection on Halting Decide?\par}
		\bigskip
		{\large Saman Farhat and Farzad Didehvar\par}
		\medskip
		{\small Amirkabir University of Technology, Tehran, Iran\\
			\href{mailto:saman.farhat@aut.ac.ir}{saman.farhat@aut.ac.ir}, \href{mailto:didehvar@aut.ac.ir}{didehvar@aut.ac.ir}\par}
		
	\end{center}
	
	\begin{abstract}
		\noindent
		A probabilistic program can produce an answer on some runs and continue
		forever on others. Suppose it halts with positive probability on every
		input and we count only the runs that halt. Aaronson et al.\ showed that
		if the correct answer must be more than twice as likely as the wrong one,
		such programs decide exactly the problems solvable with a finite list of
		halting questions chosen in advance. They asked what happens when the
		correct answer need only be more likely than the wrong one. We prove that
		the resulting class is exactly $\Delta^0_2$: the halting questions can
		now be asked one at a time, with each answer helping to determine the
		next question. This remains true if the program must halt with probability
		at least $99\%$, or with any other prescribed rational lower bound below
		one.
		
		Our main technical result is the exact query bound when the program
		gives two guarantees. Suppose it halts with probability at least $\rho$ and, among
		the runs that halt, the correct answer has probability at least $a$.
		For rational $0<\rho<1$ and $1/2<a<1$, these programs decide exactly
		the problems solvable with at most $n$ halting questions chosen in
		advance, where $n$ is the largest integer satisfying
		$\rho(a/(1-a))^n\leq 1$. Every value of $n$ occurs, and different
		values give different classes. The proof measures how much additional halting probability a change
		of answer requires, and counts these changes as in Ershov's hierarchy. The results extend to quantum machines with
		computable gates.
		
	\end{abstract}
	
	\section{Introduction}\label{sec:intro}
	
	Suppose we want to know whether a program halts on a given input.
	Toss two fair coins. On two tails, output $0$. Otherwise run the
	program and output $1$ if it halts. If the program runs forever, every
	run that produces an answer says $0$. If it halts, three quarters of
	the runs say $1$. The correct answer wins among the runs that halt,
	and the machine never has to recognize a program that runs forever.
	
	We call this rule \emph{postselection on halting}. It lets a machine
	decide the \emph{halting problem}, $\HALT$, the set of program--input pairs
	for which the program eventually stops. What else can it decide? The
	answers turn out to be classical classes of computability theory, and
	the proofs count how often a tentative answer can change, as in
	Ershov's hierarchy.
	
	Aaronson et al.\ showed that requiring the correct bit to be more
	than twice as likely as the wrong one gives exactly the languages
	\emph{truth-table reducible} to $\HALT$. For each input, an algorithm
	prepares a finite list of halting questions and a rule for combining
	the answers. Such a list is \emph{nonadaptive}: every question is fixed
	before any answer is seen. Immediately after Proposition~9.2, the authors ask what
	happens when the correct bit need only be more likely than the wrong
	one~\citep[Section~9]{AaronsonEtAl2024}.
	
	An algorithm with an oracle for $\HALT$ may also use each answer
	to choose its next question; such questions are \emph{adaptive}. A
	language decided by such an algorithm is \emph{Turing reducible} to $\HALT$. We write $\Delta^0_2$ for the
	class of these languages.
	
	\begin{introresult}[Strict majority]\label{intro:thm:A}
		Postselection on halting with a strict majority decides exactly the
		languages Turing reducible to the halting problem. The same class is
		obtained if, on every input, the machine must halt with probability
		at least $\rho$, for any prescribed rational $0<\rho<1$.
	\end{introresult}
	
	The increase in power is strict. Some languages Turing reducible to
	$\HALT$ have no truth-table reduction to it
	(Proposition~\ref{post:thm:margin}). Theorem~\ref{intro:thm:B} gives
	the exact bound on halting questions chosen in advance when we prescribe
	lower bounds on the probability of halting and on the probability of a
	correct answer among the runs that halt. The classification extends to
	quantum machines. Section~\ref{sec:discussion} summarizes its extension
	to machines that query other machines, which the first version of our
	preprint~\citep{FarhatDidehvar2026v1} carries out.
	
	\subsection{What a change of answer costs}
	
	To see why the whole of $\Delta^0_2$ is within reach, consider the
	answers visible after $t$ steps. Their probabilities can be computed
	by following all coin sequences for that many steps. Use the bit
	currently ahead as a tentative answer, resolving a tie in favor of
	$0$. If one bit eventually has strictly greater probability, this
	guess must settle on it, although we may never know when the last
	change has occurred. Shoenfield's limit lemma identifies precisely the
	languages that admit such a computable sequence of guesses: they are
	the languages in $\Delta^0_2$~\citep{Shoenfield1959}.
	
	The other direction explains how postselection can use every such
	sequence. Aaronson et al.'s construction guesses answers to a finite
	list of queries and waits for evidence for the positive guesses. We
	instead let the coins choose which change of a single guess to wait
	for. One branch reports the initial bit. Another waits for the first
	change and reports the new bit, another for the second change, and so
	on. A branch assigned to a change that never occurs waits forever.
	We choose the branch probabilities so that each change reverses the
	lead and leaves a gap half as large as before. Any finite number of
	changes can then be accommodated, even when no computable function
	bounds their number. This gives a uniform fair-coin construction from
	the program for the guesses (Theorem~\ref{post:thm:main}).
	
	A prescribed confidence changes this calculation. Suppose the machine
	must halt with probability at least $\rho$ on each input and the
	correct bit must account for at least a fraction $a>1/2$ of its
	halting probability. Call $\rho$ the \emph{halting floor} and
	$\eta=a-1/2$ the \emph{confidence margin}. To see how they interact,
	take $\rho=1/4$ and $a=2/3$. Start by allocating mass $1/6$ to
	output $0$ and mass $1/12$ to output $1$, a total of $1/4$ in which
	$0$ has a two-thirds share. For output $1$ to acquire a two-thirds
	share while the mass already allocated to $0$ is kept, that $1/6$
	must account for at most one third of the total. The total must therefore reach at least $1/2$. A second
	reversal requires a total of at least $1$. No probability remains
	for a third.
	
	\begin{introresult}[Exact query bound]\label{intro:thm:B}
		Fix rational numbers $0<\rho<1$ and $0<\eta<1/2$, and put
		$a=1/2+\eta$. Machines with halting floor $\rho$ and confidence
		margin $\eta$ decide exactly the languages truth-table reducible to
		$\HALT$ with at most $n(\rho,\eta)$ queries on each input, where
		\[
		n(\rho,\eta)=\max\left\{j\geq 0 :
		j\text{ is an integer and }\rho
		\left(\frac{a}{1-a}\right)^j\leq 1\right\}.
		\]
	\end{introresult}
	
	The bound is attainable because every increase in halting probability
	can go to the new guess. If the old guess holds a fraction $a$ of the
	total, increasing the total by the factor $a/(1-a)$ in this way gives
	the new guess precisely the same share. For the upper bound, the proof
	waits for enough evidence before adopting a guess, so small fluctuations
	in the observed majority do not use up any of the allowed changes
	(Theorem~\ref{post:thm:sharp}).
	
	Why does counting changes give a query bound? If a guess can change
	at most twice, prepare two programs. One halts when the guess first
	changes, and the other halts when it changes a second time. Asking
	$\HALT$ about both programs tells us whether to keep or reverse the
	initial guess. The same argument works for any finite bound.
	Conversely, when simulating a fixed list of halting queries, each
	positive answer is discovered only once. This connects the exact
	bound to Ershov's hierarchy of approximations by mind changes
	\citep{Ershov1968I,Ershov1968II,Ershov1970}. We allow the initial
	guess to depend computably on the input, so that it costs no query.
	For a fixed bound $n$, this correspondence between changes and queries
	is classical~\citep[Theorem~3.10]{Arslanov2011}; Theorem~\ref{intro:thm:B}
	determines $n$ from the halting floor and the confidence margin.
	Carstens's characterization at the level of a computable change
	bound gives the full truth-table class
	\citep{Carstens1976,DowneyGreenberg2018}.
	
	Table~\ref{post:tab:introladder} collects the consequences. A
	computable floor or margin may depend on the input, but an algorithm
	must give its positive rational value. Computing the halting
	probability itself means approximating its value to any requested
	accuracy. This distinction matters even for a floor arbitrarily
	close to one: knowing that at least $99$ percent of the runs halt leaves the full
	strict-majority class intact, whereas being able to compute the
	halting probability makes the language decidable.
	
	\begin{table}[htbp]
		\centering\small
		\caption{What the halting and confidence guarantees let an algorithm
			ask about $\HALT$. A fixed guarantee applies to every input. All
			positive margins in the table are strictly below $1/2$, except in
			the separate zero-error row, where every run that halts must give
			the correct answer.}
		\label{post:tab:introladder}
		\begin{tabularx}{\linewidth}{@{}
				>{\raggedright\arraybackslash}p{.28\linewidth}
				>{\raggedright\arraybackslash}p{.25\linewidth}
				>{\raggedright\arraybackslash}X@{}}
			\toprule
			Halting guarantee & Confidence guarantee & Query power \\
			\midrule
			Computable halting probability & Strict majority & Decidable \\
			Positive probability & Zero error & Decidable \\
			Prescribed floor $\rho$ & Prescribed margin $\eta$ &
			At most $n(\rho,\eta)$ nonadaptive queries \\
			Some fixed floor & Some fixed margin &
			A fixed bound on nonadaptive queries \\
			Positive probability or a computable floor & Fixed or computable margin &
			A finite nonadaptive list computed from the input \\
			Prescribed floor $\rho<1$ & Computable margin &
			A finite nonadaptive list computed from the input \\
			Positive probability or a prescribed floor $\rho<1$ & Strict majority &
			Adaptive queries ($\Delta^0_2$) \\
			\bottomrule
		\end{tabularx}
	\end{table}
	
	\subsection{Origins and extensions}\label{sec:intro-ctc}
	
	Classical work on what a probabilistic program can compute goes back to
	de Leeuw, Moore, Shannon and Shapiro~\citep{DeLeeuwEtAl1956}.
	Aaronson et al.'s question arose from two proposals for computation
	with a \emph{closed timelike curve}, a path in spacetime that returns to
	its own past. In Deutsch's model the state on the loop must be
	invariant under the computation that acts on it~\citep{Deutsch1991}.
	The postselection proposal, introduced independently by Bennett and
	Schumacher and by Svetlichny and developed by Lloyd et al., conditions
	on a specified measurement outcome
	\citep[Section~9]{AaronsonEtAl2024}\,\citep{LloydEtAl2011}.
	In polynomial time, quantum computation with postselection gives
	$\mathrm{PP}$, the class decided by a probabilistic polynomial-time
	machine with a strict acceptance threshold of one half
	\citep{Aaronson2005}. Deutschian computation with polynomially many
	bits or qubits on the loop gives polynomial space~\citep{AaronsonWatrous2009}.
	
	For an unrestricted Deutschian loop, Aaronson et al.\ obtained
	$\Delta^0_2$ for both classical and quantum machines, even with
	bounded error~\citep[Theorems~3.2 and~3.3]{AaronsonEtAl2024}.
	Theorem~\ref{intro:thm:A} gives the same class under postselection on
	halting with a strict majority. A fixed margin separates the rules.
	With positive halting probability on every input and a fixed margin
	$0<\eta<1/2$, postselection on halting gives exactly
	$\mathrm{tt}(\HALT)$, while Deutschian computation retains the full
	class. Theorem~\ref{intro:thm:B}
	refines the earlier bounded-error result by determining the exact
	query count once the halting floor is also prescribed.
	
	A separate restriction is the number of fair coins a run may toss.
	Section~\ref{sec:fewcoins} characterizes polynomially many tosses by
	a bound of $2^{p(|x|)}$ mind changes for some polynomial $p$. Section~\ref{sec:quantum} proves
	that the classical and quantum models realize the same pairs of output
	probabilities. The proof uses the classical realization of probabilities
	approximable from below~\citep{ZvonkinLevin1970}, so the query bound
	carries over even when equality holds in the condition defining $n$.
	Section~\ref{sec:discussion} compares the two rules and ends with a
	question about machines that query other machines: what can they decide
	when each query returns a fresh sample of the queried machine's halting
	output?
	
	\section{Background}\label{sec:background}
	
	Fix an effective enumeration $(P_e)_{e\geq0}$ of programs, and identify
	binary strings with the nonnegative integers in length-lexicographic
	order. Write $\one_L$ for the characteristic function of a language $L$.
	The halting problem $\HALT=\{\langle e,w\rangle:P_e(w)\text{ halts}\}$ is
	undecidable~\citep[Theorem~5.1]{Sipser2013}.
	
	\paragraph{Reductions.}
	An oracle for a language $A$ answers a membership query $y$ with the
	bit $\one_A(y)$. We write $L\le_{\mathrm T}A$ if a machine with oracle
	$A$ halts on every input and decides $L$. In a truth-table reduction
	$L\le_{\mathrm{tt}}A$, a total computable procedure produces from $x$
	a list $y_1,\ldots,y_m$ and a Boolean table $\tau_x$ such that
	\[
	\one_L(x)=\tau_x\bigl(\one_A(y_1),\ldots,\one_A(y_m)\bigr).
	\]
	Both are fixed before any oracle answers arrive. Write $\mathrm{tt}(A)$
	for the truth-table class and $\mathrm{btt}(A)$ for its subclass with a uniform
	finite query bound. The least such bound for a reduction is its
	\emph{norm}. In this notation $\Delta^0_2=\{L:L\le_{\mathrm T}\HALT\}$.
	
	For $\HALT$, this total-table convention gives the nonadaptive class
	of \citet[Proposition~9.2]{AaronsonEtAl2024}, even if their final
	computation is required to halt only on the true answer vector.
	For each candidate vector, append a halting query testing whether that
	final computation outputs $1$. A total table uses the original answers
	to select the appropriate extra bit. This conversion can increase the
	query count.
	
	\paragraph{Computable approximations.}
	A computable approximation to $L$ is a total computable Boolean function
	$g(x,t)$ with $\lim_t g(x,t)=\one_L(x)$ on every input. Shoenfield's
	limit lemma characterizes $\Delta^0_2$ by the existence of such an
	approximation~\citep{Shoenfield1959}. Its number of \emph{mind changes}
	on $x$ is $\nchg(x)=|\{t\geq1:g(x,t)\ne g(x,t-1)\}|$.
	
	In Ershov's hierarchy, a $k$-c.e.\ approximation starts at $0$ and
	changes at most $k$ times on each input~\citep{Ershov1968I,Ershov1968II,Ershov1970}.
	We allow the computable initial bit $g(x,0)$ to be either value. This
	\emph{free start} makes the uniform bound on changes agree exactly with
	the bound on queries (Lemma~\ref{post:lem:bridge}). Prefixing a zero
	costs at most one change, so the conventions have the same union over
	finite bounds and the same $\omega$-c.e.\ class, where the bound is a
	computable function of $x$. That last class is $\mathrm{tt}(\HALT)$
	\citep{Carstens1976,DowneyGreenberg2018}.
	
	A real is \emph{left-computably enumerable} if it is the limit of a
	computable nondecreasing sequence of rationals. For the machines in
	Section~\ref{sec:postselection}, the probability $p_i(x)$ of eventually
	outputting bit $i$ on input $x$ is such a real. The answer on $x$ depends
	on the sign of $p_1(x)-p_0(x)$. For the computability properties of
	differences of left-computably enumerable reals, see
	\citep{AWZ2000,Zheng2000}.
	
	\section{Unbounded error: the class \texorpdfstring{$\Delta^0_2$}{Δ⁰₂}}\label{sec:postselection}
	
	\subsection{The machine and its answer rule}
	
	\begin{definition}[Postselection on halting]\label{post:def:model}
		Let $M$ be a probabilistic Turing machine with a finite program and
		no oracle. At each transition, $M$ either takes a deterministic step
		or tosses one fresh, independent fair coin to choose between two
		instructions. On input $x\in\{0,1\}^*$, a run either halts with
		output $0$ or $1$, or continues forever.
		
		For $i\in\{0,1\}$, let $p_i(x)$ denote the probability that $M(x)$
		eventually halts with output $i$. The probability of halting is
		\[
		\Phalt(x)=p_0(x)+p_1(x)\leq 1.
		\]
		A language $L$ belongs to $\PostU$ if there is such a machine $M$
		for which, on every input $x$,
		\begin{equation}\label{post:eq:strict}
			\begin{aligned}
				x\in L    &\implies p_1(x)>p_0(x),\\
				x\notin L &\implies p_0(x)>p_1(x).
			\end{aligned}
		\end{equation}
	\end{definition}
	
	The strict inequalities in \eqref{post:eq:strict} guarantee that
	$\Phalt(x)>0$ on every input. We can therefore define the
	\emph{conditional correctness} of $M$ on $x$ by
	\[
	q(x)=\Prob[M(x)=\one_L(x)\mid M(x)\text{ halts}]
	=\frac{p_{\one_L(x)}(x)}{\Phalt(x)}.
	\]
	The decision rule is $q(x)>1/2$. We call $q(x)-1/2$ the
	\emph{conditional advantage}. Running time, space and the number of
	coins are unrestricted. The missing probability $1-\Phalt(x)$ is the
	probability of divergence~\citep[Section~2]{DLZG2014}.
	
	If the correct output had \emph{unconditional} probability greater
	than $1/2$, a decider could explore the finite computation trees until
	one output accumulated mass above $1/2$. Almost-sure halting with a
	separate discard output also gives decidability
	\citep[Proposition~9.1]{AaronsonEtAl2024}.
	
	\subsection{The characterization}
	
	The construction adapts the guess-and-diverge argument of
	\citet[proof of Proposition~9.2]{AaronsonEtAl2024}. Here the random
	choice specifies which change in a computable approximation to wait for.
	
	\begin{theorem}[Strict-majority classification]\label{post:thm:main}
		The class $\PostU$ equals $\Delta^0_2$. For every fixed rational
		$\rho$ with $0<\rho<1$, requiring $\Phalt(x)\geq\rho$ on every input
		still gives exactly $\Delta^0_2$. Requiring $\Phalt(x)=1$ on every
		input gives exactly the decidable languages, as already follows from
		\citep[Proposition~9.1]{AaronsonEtAl2024}.
		
		For $0<\rho<1$, the constructions in both directions are effective.
		Given the code of a machine witnessing $L\in\PostU$, one can construct
		a program for a computable approximation to $L$. Given the code of
		such an approximation and the rational number $\rho$, one can
		construct a machine witnessing $L\in\PostU$ with $\Phalt(x)\geq\rho$
		on every input.
	\end{theorem}
	
	\begin{proof}
		\emph{Upper bound.} Let $p_{i,t}(x)$ be the probability of halting with
		output $i$ within $t$ transitions. It is a computable dyadic rational:
		simulate $t$ transitions on all $2^t$ bit strings, counting all
		completions of any prefix on which the machine halts early. Since
		$p_{i,t}(x)\uparrow p_i(x)$, the computable guess
		$\one\{p_{1,t}(x)>p_{0,t}(x)\}$ converges to $\one_L(x)$ by
		\eqref{post:eq:strict}. The limit lemma gives $L\in\Delta^0_2$.
		This is the standard limit argument for the sign of a difference of
		two left-computably enumerable reals (Section~\ref{sec:background}).
		
		\emph{Lower bound.}
		Let $L\in\Delta^0_2$, and fix a computable approximation $g$ to $L$.
		The machine first samples an integer $J\geq 0$. The value $J=0$
		selects the initial guess, and $J=k\geq 1$ selects the $k$th mind
		change. We use the probabilities
		\begin{equation}\label{post:eq:weights}
			w_0=\Prob[J=0]=\frac14,\qquad
			w_k=\Prob[J=k]=\frac{3}{2^{k+2}}\quad(k\geq 1).
		\end{equation}
		
		To sample $J$, toss two fair coins. On outcome $00$, return $J=0$.
		On any of the other three outcomes, toss fresh coins until the first
		head and return the number of fresh tosses, including the head.
		The first head occurs on toss $k$ with probability $2^{-k}$, so the
		overall probability of returning $k\geq 1$ is $(3/4)2^{-k}=w_k$.
		These weights, together with $w_0$, sum to one. If every fresh toss
		is a tail, the sampler runs forever. This event has probability zero.
		
		If $J=0$, the machine outputs the initial guess $g(x,0)$. If
		$J=k\geq 1$, it computes $g(x,0),g(x,1),\ldots$ in order. At the
		$k$th change, it outputs the new guess. Since $g$ is total, the
		machine reaches every finite stage for which it is still waiting.
		If the $k$th change never occurs, the machine runs forever.
		
		Fix an input $x$, and let $m=\nchg(x)$ be the total number of mind
		changes. This number is finite because $g(x,t)$ eventually stabilizes.
		Exactly the branches with $J=0,\ldots,m$ halt, so
		\begin{equation}\label{post:eq:haltmass}
			\Phalt(x)
			=\frac14+\sum_{k=1}^{m}\frac{3}{2^{k+2}}
			=1-\frac{3}{2^{m+2}}
			\geq\frac14.
		\end{equation}
		
		To see which output has greater probability, write
		$d(x)=p_1(x)-p_0(x)$ and let $\sigma_0=2g(x,0)-1$ be the sign of
		the initial guess. The initial branch contributes $\sigma_0/4$ to
		$d(x)$. Each mind change flips the sign. After the first $k$ changes,
		the signed sum of the contributions is
		$(-1)^k\sigma_0/2^{k+2}$. Indeed, this holds for $k=0$. At change
		$k+1$, the new branch contributes $3/2^{k+3}$ with the opposite
		sign. Since the previous sum has magnitude $2/2^{k+3}$, the new sum
		has the sign of the new guess and magnitude $1/2^{k+3}$.
		
		After all $m$ changes, the guess is $\one_L(x)$. Thus
		\begin{equation}\label{post:eq:sign}
			d(x)=\frac{(-1)^m\sigma_0}{2^{m+2}}
			=\frac{2\one_L(x)-1}{2^{m+2}}.
		\end{equation}
		This gives $p_1(x)>p_0(x)$ on yes-inputs and $p_0(x)>p_1(x)$ on
		no-inputs, including when the initial guess never changes.
		
		\emph{A prescribed halting floor.} Given rational $0<\rho<1$, choose
		$c=2^{-k}$ with $k\geq0$ and $3c/4\leq1-\rho$. Run the machine just
		constructed with probability $c$ and otherwise halt with a fair output
		bit. The new signed gap is $cd(x)$ and the new halting probability is
		$1-c+c\Phalt(x)\geq1-3c/4\geq\rho$. The code is computable from $g$
		and $\rho$ throughout. At $\rho=1$ the unconditional correct-output
		probability exceeds $1/2$, so finite-tree search decides the language.
		Conversely, any deterministic decider meets this endpoint.
	\end{proof}
	
	For the original machine, before we add the fair output bit, the
	conditional advantage has an exact expression in the number of
	changes, which need not be computable from $x$. If the approximation
	changes $\nchg(x)$ times on input $x$, the correct output has
	probability $(\Phalt(x)+|d(x)|)/2$, so
	\eqref{post:eq:haltmass} and \eqref{post:eq:sign} give
	\begin{equation}\label{post:eq:advantage}
		q(x)-\frac12
		=\frac{|d(x)|}{2\Phalt(x)}
		=\frac{1}{2\bigl(2^{\nchg(x)+2}-3\bigr)}.
	\end{equation}
	Each additional mind change reverses the sign of the gap $d(x)$ and
	halves its magnitude. The halting probability stays at least $1/4$,
	but the conditional advantage may have no computable positive lower
	bound. Indeed, by \eqref{post:eq:advantage} the advantage decreases
	strictly as $\nchg(x)$ grows, so a computable positive lower bound on
	it would give a computable upper bound on $\nchg(x)$. A language with
	such a bound is $\omega$-c.e.\ and hence truth-table reducible to
	$\HALT$ (Section~\ref{sec:background}), so for the languages outside
	$\mathrm{tt}(\HALT)$ given by Proposition~\ref{post:thm:margin} no
	such lower bound exists.
	
	\subsection{A worked example}\label{post:example}
	
	Table~\ref{post:tab:changes} follows the construction from the initial
	guess $0$. The first reversal gives output $1$ a lead of $1/8$, and the
	second gives output $0$ a lead of $1/16$.
	
	\begin{table}[htbp]
		\centering
		\caption{Output probabilities for an approximation starting at $0$.
			Each row assumes exactly $m$ mind changes. The final guess is the
			correct bit, and $q$ is its probability conditional on halting.}
		\label{post:tab:changes}
		\begin{tabular}{@{}cccccc@{}}
			\toprule
			Changes $m$ & Final guess & $p_0$ & $p_1$ & $\Phalt$ & $q$ \\
			\midrule
			$0$ & $0$ & $1/4$  & $0$     & $1/4$   & $1$     \\
			$1$ & $1$ & $1/4$  & $3/8$   & $5/8$   & $3/5$   \\
			$2$ & $0$ & $7/16$ & $3/8$   & $13/16$ & $7/13$  \\
			$3$ & $1$ & $7/16$ & $15/32$ & $29/32$ & $15/29$ \\
			\bottomrule
		\end{tabular}
	\end{table}
	
	For $\HALT$, start at $0$ and switch to $1$ on observing a halt.
	The first two rows therefore give conditional correctness $1$ on a
	nonhalting pair and $3/5$ on a halting pair.
	
	\subsection{Separation from the truth-table class}
	
	\begin{proposition}[Separation of the two query classes]\label{post:thm:margin}
		The class $\mathrm{tt}(\HALT)$ is a proper subclass of
		$\Delta^0_2=\PostU$.
	\end{proposition}
	
	\begin{proof}
		The equality follows from Theorem~\ref{post:thm:main}. For strictness,
		use $\HALT$ to decide whether $P_e(e)$ halts. If it diverges or returns
		an invalid truth-table description, set $A(e)=0$. Otherwise evaluate
		its table using $\HALT$ and take the opposite bit. Thus $A$ is Turing
		reducible to $\HALT$, and a program of index $e$ computing a truth-table
		reduction for $A$ would give the wrong answer on $e$.
	\end{proof}
	
	\section{The exact floor--confidence trade-off}\label{sec:ladder}
	
	We now prescribe lower bounds on the probability of halting and on
	confidence conditional on halting. Their interaction determines exactly how often
	a computable approximation can change its answer.
	
	\subsection{Two scales of guarantees}
	
	\begin{definition}[Halting guarantees]\label{ladder:def:halting}
		Let $M$ be a machine as in Definition~\ref{post:def:model}, and fix
		a rational number $0<\rho<1$. We use the following labels for
		conditions on its halting probability $\Phalt$.
		\begin{itemize}
			\item[(H0)] \emph{Positive halting probability.} On every input $x$,
			$\Phalt(x)>0$.
			\item[(H$\rho$)] \emph{Fixed halting floor.} On every input $x$,
			$\Phalt(x)\geq\rho$. The same floor $\rho$ applies to all inputs.
			\item[(Hc)] \emph{Computable halting probability.} A single algorithm,
			given any input $x$ and precision parameter $k\in\N$, halts and
			returns a rational number $r$ with $|r-\Phalt(x)|\leq 2^{-k}$.
			\item[(H1)] \emph{Almost-sure halting.} On every input $x$,
			$\Phalt(x)=1$.
		\end{itemize}
		We also allow the floor to depend on the input. A \emph{computable
			floor} is a total computable function $\rho(x)$ with rational values
		in $(0,1)$ such that $\Phalt(x)\geq\rho(x)$ for every $x$.
	\end{definition}
	
	\begin{definition}[Confidence guarantees]\label{ladder:def:confidence}
		Fix a language $L$ and a machine $M$ with positive halting probability
		on every input. Let $q(x)$ be the probability that $M$ outputs
		$\one_L(x)$ conditional on halting on input $x$.
		\begin{itemize}
			\item[(C0)] \emph{Strict majority.} On every input $x$, $q(x)>1/2$.
			\item[(C$\eta$)] \emph{Fixed confidence margin.} For a prescribed
			rational number $0<\eta\leq 1/2$, we require
			$q(x)\geq 1/2+\eta$ on every input $x$. We call $\eta$ the
			\emph{confidence margin} and $a=1/2+\eta$ the \emph{confidence
				threshold}. The endpoint $\eta=1/2$ gives \emph{zero error}, since
			then $q(x)=1$ on every input.
			\item[(Ceff)] \emph{Computable margin.} There is a total computable
			function $\eta(x)$ with rational values in $(0,1/2)$ such that
			$q(x)\geq 1/2+\eta(x)$ on every input $x$.
		\end{itemize}
	\end{definition}
	
	A machine that halts almost surely has the constant halting probability
	$\Phalt(x)=1$, so (H1) implies (Hc). Both (H1) and any positive
	halting floor imply (H0). Every fixed positive confidence margin
	implies (Ceff), and (Ceff) implies (C0). At the zero-error endpoint,
	for example, the constant function $\eta(x)=1/4$ witnesses (Ceff).
	A deterministic decider satisfies (H1) with zero error.
	
	Throughout this section, $p_{i,t}(x)$ denotes the probability that
	the machine halts with output $i$ within $t$ transitions. As shown
	in the proof of Theorem~\ref{post:thm:main}, this is an exactly
	computable dyadic rational. Write
	$P_t(x)=p_{0,t}(x)+p_{1,t}(x)$ for the probability of halting
	within $t$ transitions. As $t$ increases, $p_{i,t}(x)$ and $P_t(x)$
	converge from below to $p_i(x)$ and $\Phalt(x)$, respectively.
	
	\Needspace{8\baselineskip}
	\subsection{Computable halting probability and zero error}
	
	\begin{proposition}\label{post:prop:haltcomputable}
		Under (C0) and (Hc), the machines decide exactly the decidable
		languages. Given the code of a machine $M$ and a program witnessing
		(Hc), one can construct a decider for its language. The converse
		construction is also effective.
	\end{proposition}
	
	\begin{proof}
		From the witness for (Hc), compute $r_t$ within $2^{-t}$ of $\Phalt(x)$.
		For each bit $i$,
		\[
		p_{i,t}(x)\le p_i(x)\le r_t+2^{-t}-p_{1-i,t}(x).
		\]
		These computable intervals have width $r_t+2^{-t}-P_t(x)\to0$.
		By (C0), the limits $p_0(x)$ and $p_1(x)$ differ. Search until the
		two intervals are disjoint and return the bit with the larger lower
		endpoint. Conversely, a deterministic decider has $q=\Phalt=1$.
	\end{proof}
	
	\begin{proposition}[Zero error]\label{post:prop:zeroerror}
		Under (H0) and zero error, the machines decide exactly the decidable
		languages.
	\end{proposition}
	
	\begin{proof}
		Every halting run has a finite coin prefix of positive probability.
		Under zero error its answer is correct. By (H0), a search through the
		finite computation trees eventually finds such a prefix. Conversely,
		a deterministic decider meets both requirements.
	\end{proof}
	
	\subsection{Mind changes, queries and masses}
	
	For $n\geq1$, a set is called \emph{weakly $n$-c.e.} if it has a
	computable free-start approximation that makes at most $n$ changes.
	Such an approximation may start with either bit.
	Arslanov~\citep[Theorem~3.10]{Arslanov2011} credits their
	characterization by $n$ nonadaptive halting queries to Epstein, Haas
	and Kramer~\citep{EpsteinHaasKramer1981} and to
	Carstens~\citep{Carstens1976}. We give the effective conversions
	below. Theorem~\ref{post:thm:sharp} determines $n$ from the two
	probability guarantees.
	
	\begin{lemma}\label{post:lem:bridge}
		Let $L$ be a language and let $n\geq 0$ be an integer.
		\begin{enumerate}
			\item There is a truth-table reduction from $L$ to $\HALT$ that
			uses at most $n$ queries on every input if and only if $L$ has a
			computable free-start approximation that makes at most $n$ mind
			changes on every input.
			\item $L\le_{\mathrm{tt}}\HALT$ if and only if $L$ has a computable
			free-start approximation for which a total computable function
			$h(x)$ bounds the number of mind changes on input $x$.
		\end{enumerate}
		The constructions are effective. From a truth-table reduction one
		can compute an approximation and a bound on its mind changes.
		Conversely, from an approximation and the bound $n$ in part~(1),
		or a program for the bound $h$ in part~(2), one can compute the
		corresponding reduction.
	\end{lemma}
	
	\begin{proof}
		Given $m$ queries and a table $\tau_x$, evaluate the table at stage $t$
		using the bits observed by simulating the queried computations for $t$
		steps. Each bit turns to $1$ at most once. The table therefore changes
		value at most $m$ times and eventually gives $\one_L(x)$.
		
		Conversely, given an approximation $g$ and a bound $n$, prepare one
		query for each $j=1,\ldots,n$: does the program that watches $g(x,t)$
		ever witness its $j$th change? The final bit is the exclusive or of
		$g(x,0)$ and all the answers, also when the query list is empty.
		For part~(2), the first construction supplies the computable bound
		$m(x)$, and the second works input by input with $n=h(x)$ for the
		supplied computable bound $h$. Both transformations are uniform.
	\end{proof}
	
	Our second tool lets us construct a machine by specifying how much
	probability it should assign to each output. A \emph{mass program}
	consists of two total computable functions $\ell_0(x,t)$ and
	$\ell_1(x,t)$ with nonnegative rational values. For each input $x$,
	both functions are nondecreasing in the stage $t$, and at every stage
	their sum is at most $1$. Thus, for $i\in\{0,1\}$, the limit
	\[
	p_i(x)=\lim_{t\to\infty}\ell_i(x,t)
	\]
	exists and is left-computably enumerable uniformly in $x$
	(Section~\ref{sec:background}).
	
	\begin{lemma}[Exact realization of output masses]\label{post:lem:realize}
		Given a mass program, one can compute the code of a machine $M$ as
		in Definition~\ref{post:def:model} such that, for every input $x$
		and each $i\in\{0,1\}$,
		\[
		\Prob[M(x)\text{ eventually halts with output }i]=p_i(x).
		\]
	\end{lemma}
	
	\begin{proof}
		Fix an input $x$. At stage $t=1,2,\ldots$, the machine reads one
		fresh fair coin bit. Let $k$ be the integer represented by the first
		$t$ bits, including any leading zeros. These bits select the interval
		\[
		D_t=[k2^{-t},(k+1)2^{-t}).
		\]
		The machine computes $\ell_0(x,t)$ and $\ell_1(x,t)$. It outputs $0$
		and halts if $D_t\subseteq[0,\ell_0(x,t))$, and it outputs $1$ and
		halts if $D_t\subseteq[1-\ell_1(x,t),1)$. Otherwise it proceeds to
		stage $t+1$. The two output intervals are disjoint because
		$\ell_0(x,t)+\ell_1(x,t)\leq 1$. All endpoint comparisons are
		between rationals, and the machine reads its coins one at a time.
		
		To compute the output probabilities, consider all $2^t$ possible
		prefixes of length $t$, including extensions of prefixes on which
		the machine has already halted. Each prefix has probability $2^{-t}$.
		As $t$ increases, each output interval expands or stays fixed, and
		each new bit selects a subinterval of the preceding $D_t$. Thus a
		prefix gives output $i$ by stage $t$ exactly when its interval $D_t$
		lies inside the stage-$t$ interval for output $i$.
		
		The interval for output $0$ starts at $0$, and the interval for
		output $1$ ends at $1$. Each therefore contains exactly
		$\lfloor 2^t\ell_i(x,t)\rfloor$ whole intervals of length $2^{-t}$
		from the dyadic partition. Hence
		\[
		\Prob[M(x)\text{ halts with output }i\text{ by stage }t]
		=2^{-t}\lfloor 2^t\ell_i(x,t)\rfloor.
		\]
		This probability differs from $\ell_i(x,t)$ by less than $2^{-t}$,
		so it converges to $p_i(x)$. Since every halting run finishes at
		some finite stage, the probability of eventually outputting $i$ is
		exactly $p_i(x)$.
		
		The construction is uniform in the programs for $\ell_0$ and
		$\ell_1$.
	\end{proof}
	
	This lemma is the two-output case of a classical characterization
	going back to Zvonkin and Levin~\citep{ZvonkinLevin1970}: the output
	probabilities of probabilistic machines are exactly the uniformly
	left-computably enumerable families of masses with total at most one.
	See also \citep{Shen2015Around}. We use the lemma for each lower
	bound below and again in Section~\ref{sec:quantum} to transfer the
	results to quantum machines.
	
	\subsection{The sharp halting-floor theorem}
	
	Fix a rational halting floor $\rho\in(0,1)$ and a rational confidence
	margin $\eta\in(0,1/2)$. Write $a=\tfrac12+\eta$ for the required
	confidence and $R=a/(1-a)$ for the corresponding ratio of correct
	to incorrect output mass. Equivalently, $R=(1+2\eta)/(1-2\eta)$.
	Define
	\[
	n(\rho,\eta)=\max\{j\in\N : \rho R^j\leq 1\},
	\]
	where the maximum includes $j=0$. Since $\rho<1$ and $R>1$, it is
	well defined. Starting at $\rho$ and multiplying by $R$ until the
	product first exceeds $1$ computes $n(\rho,\eta)$ exactly. Equivalently,
	\[
	n(\rho,\eta)=\floor{\frac{\log(1/\rho)}{\log R}}.
	\]
	The rational test includes the boundary case $\rho R^j=1$.
	
	\begin{theorem}[Sharp halting-floor theorem]\label{post:thm:sharp}
		Let $n=n(\rho,\eta)$. A language $L$ is decided by a machine with
		halting floor $\rho$ and conditional correctness at least
		$a=\tfrac12+\eta$ on every input if and only if $L$ is truth-table
		reducible to $\HALT$ with at most $n$ queries on every input.
		Given $\rho$ and $\eta$, the transformations from a machine to a
		reduction and from a reduction to a machine are both effective.
	\end{theorem}
	
	\begin{proof}
		\emph{Upper bound.}
		Suppose $M$ decides $L$ under (H$\rho$) and (C$\eta$). By
		Lemma~\ref{post:lem:bridge}(1), it suffices to construct a computable
		approximation to $L$ that makes at most $n$ mind changes on every
		input.
		
		The guarantees on $M$ concern the limiting output masses. To
		recognize evidence for an answer at a finite stage, we use slightly
		smaller thresholds $\rho_0<\rho$ and $a_0<a$. We can choose them
		so that
		\[
		\rho_0R_0^{\,n+1}>1,
		\qquad R_0=\frac{a_0}{1-a_0}.
		\]
		Indeed, $\rho R^{n+1}>1$ by the definition of $n$. For
		$m=1,2,\ldots$, try the rational values
		\[
		\rho_0=\rho(1-2^{-m}),
		\qquad a_0=a-2^{-m}\eta.
		\]
		They increase to $\rho$ and $a$, respectively, so the displayed
		strict inequality eventually holds. This search computes suitable
		thresholds with $\rho_0>0$ and $a_0>1/2$.
		
		Fix an input $x$. For $i\in\{0,1\}$, call stage $s$ \emph{$i$-good}
		if
		\[
		P_s(x)\geq\rho_0
		\qquad\text{and}\qquad
		p_{i,s}(x)\geq a_0P_s(x).
		\]
		Both conditions can be checked at stage $s$. Since $\rho_0>0$ and
		$a_0>1/2$, at most one output can meet them at a given stage.
		
		Let $b=\one_L(x)$ be the correct bit, and suppress $x$ in the
		probabilities below. Every sufficiently late stage is $b$-good.
		Indeed, $P_s\to\Phalt\geq\rho>\rho_0$, and
		\[
		p_{b,s}-a_0P_s
		\longrightarrow p_b-a_0\Phalt
		\geq (a-a_0)\Phalt>0.
		\]
		We can therefore search for the first stage $s_0$ that is good for
		either bit. Set $g(x,0)$ to the bit found at that stage. For $t\geq 1$,
		if stage $s_0+t$ is good for some bit, set $g(x,t)$ to that bit.
		Otherwise, keep the previous guess. The search for $s_0$ terminates,
		and all subsequent tests compare rational numbers, so $g$ is total
		and computable. Since every sufficiently late stage is $b$-good and
		no stage is good for both bits, $g(x,t)$ eventually settles on $b$.
		
		To bound the number of changes, suppose the approximation adopts
		the guess $i$ at stage $u$, either initially or at a change. Let
		$v>u$ be the stage of the next change, if there is one. At stage $u$,
		output $i$ accounts for at least a fraction $a_0$ of the observed
		halting probability. At stage $v$, the opposite output accounts for
		at least that fraction. Since the mass already assigned to $i$
		cannot decrease,
		\[
		a_0P_u
		\leq p_{i,u}
		\leq p_{i,v}
		\leq (1-a_0)P_v.
		\]
		Thus $P_v\geq R_0P_u$. At the initial stage $P_{s_0}\geq\rho_0$,
		so after $k$ changes the observed halting probability is at least
		$\rho_0R_0^k$. An $(n+1)$st change would make this probability
		greater than $1$, by our choice of $\rho_0$ and $R_0$. Hence the
		approximation makes at most $n$ changes.
		
		\emph{Lower bound.}
		Suppose $L$ has a truth-table reduction to $\HALT$ with at most
		$n$ queries on every input. By Lemma~\ref{post:lem:bridge}(1), we
		can compute a free-start approximation $g$ to $L$ with at most $n$
		mind changes. Let $k(x,t)$ be the number of changes through stage
		$t$. At that stage, assign total mass
		\[
		T_t(x)=\rho R^{k(x,t)}
		\]
		and divide it between the two outputs according to the current guess:
		\[
		\ell_{g(x,t)}(x,t)=aT_t(x),
		\qquad
		\ell_{1-g(x,t)}(x,t)=(1-a)T_t(x).
		\]
		These masses are computable rationals, and their sum is at most
		$\rho R^n\leq 1$.
		
		It remains to check that a change of guess never requires us to
		withdraw mass from either output. Suppose the total just before a
		change is $T$. After the change it is $RT$. The former guess had
		mass $aT$ and now has mass $(1-a)RT=aT$. Its mass therefore stays
		fixed, and the entire increase $(R-1)T$ goes to the new guess.
		Between changes both masses stay fixed. Thus $(\ell_0,\ell_1)$ is
		a mass program, and Lemma~\ref{post:lem:realize} gives a machine
		with exactly these limiting output probabilities.
		
		If $g$ makes $k$ changes on $x$, the machine has halting probability
		$\rho R^k\geq\rho$. The final guess is $\one_L(x)$ and receives
		exactly the fraction $a$ of that probability. Hence the machine
		satisfies both (H$\rho$) and (C$\eta$), including the boundary case
		$\rho R^n=1$. The threshold examples below illustrate this case.
	\end{proof}
	
	\begin{corollary}[Exact thresholds]\label{post:cor:threshold}
		Let $\rho\in(0,1)$ and $\eta\in(0,1/2)$ be rational, and put
		$a=\tfrac12+\eta$.
		\begin{enumerate}
			\item The languages decidable under (H$\rho$) and (C$\eta$) are
			exactly the decidable languages if and only if $\rho a>1-a$.
			\item At the boundary $\rho a=1-a$, we have $n(\rho,\eta)=1$.
			In particular, $\HALT$ can be decided under these guarantees.
			\item Lowering either $\rho$ or $\eta$ can only enlarge the class.
			Every query bound occurs: $n(3/4,1/6)=0$, and
			$n(2^{-j},1/6)=j$ for every $j\geq 1$. The union over all allowed
			rational pairs $(\rho,\eta)$ is $\mathrm{btt}(\HALT)$.
			\item Distinct values of $n(\rho,\eta)$ give distinct classes.
			For every $n\geq 0$, some language has a truth-table reduction to
			$\HALT$ using at most $n+1$ queries on every input, but has no such
			reduction using at most $n$.
		\end{enumerate}
	\end{corollary}
	
	\begin{proof}
		For part~(1), recall that $R=a/(1-a)$. The query bound is zero
		exactly when $\rho R>1$, or equivalently $\rho a>1-a$.
		A truth-table reduction with zero queries is a decider. If
		$\rho a\leq 1-a$, the bound is at least one, so
		Theorem~\ref{post:thm:sharp} puts the undecidable language $\HALT$
		in the class. At equality, $\rho R=1$ and $\rho R^2=R>1$, giving
		the exact bound of one in part~(2).
		
		For part~(3), lowering either guarantee preserves every machine
		that already satisfies it. To realize the query bounds, fix
		$\eta=1/6$, so that $a=2/3$ and $R=2$. Taking $\rho=3/4$ gives
		$\rho R>1$ and hence a bound of zero. For $j\geq 1$, taking
		$\rho=2^{-j}$ gives $\rho R^j=1$ and $\rho R^{j+1}=2$, so the
		bound is exactly $j$. The union claim now follows from
		Theorem~\ref{post:thm:sharp}: each pair gives a finite uniform
		query bound, and every such bound occurs.
		
		For part~(4), fix $n\geq 0$. By Lemma~\ref{post:lem:bridge}(1),
		it suffices to construct a language with a computable approximation
		that makes at most $n+1$ mind changes on each input, and to rule
		out every approximation with at most $n$ changes.
		
		Enumerate the partial computable functions of two arguments as
		$\varphi_0,\varphi_1,\ldots$, and identify inputs with natural
		numbers as in Section~\ref{sec:background}. On
		input $e$, set $G(e,0)=0$. At each stage $t\geq 1$, simulate
		$\varphi_e(e,j)$ for $t$ steps for every $j\leq t$. Look for the
		largest $s\leq t$ for which all computations with $0\leq j\leq s$
		have halted with values in $\{0,1\}$. If there is no such $s$,
		keep the previous guess. Otherwise, count the changes in the
		verified prefix
		\[
		\varphi_e(e,0),\ldots,\varphi_e(e,s).
		\]
		If the count is at most $n$, set $G(e,t)=1-\varphi_e(e,s)$.
		If the count exceeds $n$, keep the previous guess. Each stage
		requires only a finite computation.
		
		The verified prefixes extend one another. Adopting the complement
		of the first verified prefix's last bit costs at most one change.
		Each subsequent change of $G$ requires a further change in the
		verified prefix. Once a prefix contains more than $n$ changes,
		$G$ keeps its value at every later stage. Thus $G$ changes at most
		$n+1$ times on each input and has a limit.
		
		Let $A$ be the language with $\one_A(e)=\lim_t G(e,t)$. Suppose
		some total $\varphi_e$ approximates $A$ with at most $n$ changes
		on every input. On input $e$, its verified prefixes grow without
		bound, and each contains at most $n$ changes. The construction
		therefore gives
		\[
		\one_A(e)=\lim_t G(e,t)
		=1-\lim_s\varphi_e(e,s)
		=1-\one_A(e),
		\]
		a contradiction. This separates each query bound from the next.
	\end{proof}
	
	\paragraph{Threshold examples.}
	For $\rho=1/4$ and $a=2/3$, the example in the introduction attains
	two reversals. Starting at $0$, the successive pairs $(p_0,p_1)$ of
	allocated masses are
	\[
	(1/6,1/12)\;\longrightarrow\;(1/6,1/3)\;\longrightarrow\;(2/3,1/3).
	\]
	At the one-query boundary $\rho=1/2$, $a=2/3$, the initial pair
	$(1/3,1/6)$ already favors $0$. To decide $\HALT$, raise the second
	mass to $2/3$ if the simulated program halts. Either possible final
	pair gives the required confidence. At the endpoints $\rho=1$ or
	$a=1$, Propositions~\ref{post:prop:haltcomputable}
	and~\ref{post:prop:zeroerror} give decidability.
	
	\subsection{Computable guarantees}
	
	Positive halting probability on every input already gives a
	computable floor that may depend on the input. Given $x$, search
	for a stage $t$ with $P_t(x)>0$ and return $P_t(x)/2$. The search
	terminates under (H0), and the resulting rational lies in $(0,1)$
	and is at most $\Phalt(x)$. This lets us apply the query bound of
	Theorem~\ref{post:thm:sharp} separately on each input.
	
	\begin{proposition}[Computable guarantees]\label{post:prop:ttrows}
		For a language $L$, the following are equivalent.
		\begin{enumerate}
			\item $L\le_{\mathrm{tt}}\HALT$.
			\item Some machine decides $L$ under (H0) and (Ceff).
			\item For some rational $\eta\in(0,1/2)$, a machine decides $L$
			under (H0) and (C$\eta$). Equivalently, one may prescribe any
			rational $\eta$ in this range and require a computable halting floor.
			\item Some machine decides $L$ with a computable halting floor
			and a computable confidence margin~(Ceff).
			\item For some rational $\rho\in(0,1)$, a machine decides $L$
			under (H$\rho$) and (Ceff). Equivalently, one may prescribe any
			rational $\rho$ in this range.
		\end{enumerate}
		The conversions are effective given the programs witnessing the
		computable guarantees and the values of any prescribed rational
		parameters.
	\end{proposition}
	
	\begin{proof}
		We first show that (2) implies (1). Let $M$ decide $L$ under (H0)
		with computable margin $\eta(x)$. As above, on input $x$ we find
		the first stage $t$ with $P_t(x)>0$ and set $\rho(x)=P_t(x)/2$.
		The upper-bound construction of Theorem~\ref{post:thm:sharp} is
		uniform in its two rational parameters, so on input $x$ we can use
		$\rho(x)$ and $\eta(x)$. This gives a computable approximation to
		$L$ with at most $n(\rho(x),\eta(x))$ mind changes on input $x$.
		The bound is computable from $x$ by rational arithmetic.
		Lemma~\ref{post:lem:bridge}(2) now gives $L\le_{\mathrm{tt}}\HALT$.
		
		Each of (3), (4), and (5) implies (2), since a fixed margin is
		computable and either kind of positive floor implies (H0). It
		remains to obtain these guarantees from (1).
		
		Suppose $L\le_{\mathrm{tt}}\HALT$. By
		Lemma~\ref{post:lem:bridge}(2), we can compute an approximation to
		$L$ together with a function $h(x)$ bounding its number of mind
		changes. The lower-bound construction of Theorem~\ref{post:thm:sharp}
		starts with mass $\rho$ and multiplies it by $R$ at each change.
		Thus it suffices to choose the parameters on each input so that
		$\rho R^{h(x)}\leq 1$.
		
		For (3), prescribe any rational $\eta\in(0,1/2)$ and put
		$R=(1+2\eta)/(1-2\eta)$. Choose the computable floor
		\[
		\rho(x)=R^{-(h(x)+1)}.
		\]
		Then $\rho(x)R^{h(x)}=1/R<1$, including when $h(x)=0$.
		The construction therefore gives a machine with halting probability
		at least $\rho(x)$ and conditional correctness exactly
		$\tfrac12+\eta$ on every input.
		
		For (5), prescribe any rational $\rho\in(0,1)$. For each integer
		$m\geq 2$, put $\eta_m=2^{-m}$ and
		$R_m=(1+2\eta_m)/(1-2\eta_m)$. On input $x$, search for the first
		$m$ such that
		\[
		\rho R_m^{h(x)}\leq 1.
		\]
		Each test uses rational arithmetic. For fixed $x$, the left-hand
		side tends to $\rho<1$ as $m$ increases, so the search terminates.
		Set $\eta(x)=\eta_m$ for the selected $m$. The same construction
		now gives a machine with the prescribed floor $\rho$ and computable
		margin $\eta(x)$. This proves (5) and also (4), since the constant
		function $\rho(x)=\rho$ is a computable floor.
	\end{proof}
	
	Together with Proposition~\ref{post:thm:margin}, this shows that
	requiring a computable positive lower bound on the conditional advantage
	strictly reduces the class of languages decided.
	
	Aaronson et al.\ state the following result for the factor $c=2$ and
	observe that their construction gives the factor $99$
	\citep[Proposition~9.2 and its proof]{AaronsonEtAl2024}. We record the
	formulation for every fixed real $c>1$; the prescribed-margin
	construction gives it directly.
	
	\begin{corollary}[Any fixed factor]\label{post:cor:factor}
		Fix a real number $c>1$. A language belongs to $\mathrm{tt}(\HALT)$
		if and only if it is decided by a machine for which, on every input,
		the correct output is more than $c$ times as likely as the incorrect
		output.
	\end{corollary}
	
	\begin{proof}
		Put $a_c=c/(1+c)$, so that $1/2<a_c<1$. The strict factor condition
		forces the correct output to have positive probability, hence (H0)
		holds. Conditional on halting, the condition becomes
		\[
		q(x)>c\bigl(1-q(x)\bigr),
		\]
		or equivalently $q(x)>a_c$.
		
		Choose a rational $a_-$ with $1/2<a_-<a_c$. Any machine satisfying
		the factor condition has the fixed positive margin $a_--1/2$.
		Proposition~\ref{post:prop:ttrows} therefore puts its language in
		$\mathrm{tt}(\HALT)$.
		
		Conversely, let $L\in\mathrm{tt}(\HALT)$ and choose a rational
		$a_+$ with $a_c<a_+<1$. Proposition~\ref{post:prop:ttrows}(3),
		applied with margin $a_+-1/2$, gives a machine for $L$ satisfying
		(H0) and $q(x)\geq a_+$ on every input. Since $a_+>a_c$, the
		machine satisfies the strict factor condition.
	\end{proof}
	
	\subsection{The ladder}
	
	\begin{theorem}[The reducibility ladder]\label{post:thm:ladder}
		For the machines of Definition~\ref{post:def:model}, the guarantees
		in Table~\ref{post:tab:introladder} give the following classes.
		\begin{enumerate}
			\item A strict majority (C0) with computable halting probability
			(Hc) gives exactly the decidable languages. This includes (C0)
			with almost-sure halting (H1). Zero error with positive halting
			probability (H0) also gives exactly the decidable languages
			(Propositions~\ref{post:prop:haltcomputable}
			and~\ref{post:prop:zeroerror}).
			
			\item For prescribed rationals $\rho\in(0,1)$ and $\eta\in(0,1/2)$,
			the floor (H$\rho$) and margin (C$\eta$) give exactly the languages
			truth-table reducible to $\HALT$ with at most $n(\rho,\eta)$ queries
			on every input (Theorem~\ref{post:thm:sharp}). The union over all
			such pairs is $\mathrm{btt}(\HALT)$
			(Corollary~\ref{post:cor:threshold}).
			
			\item Under (H0), either a prescribed rational margin
			$\eta\in(0,1/2)$ or a computable margin (Ceff) gives exactly
			$\mathrm{tt}(\HALT)$. The same holds if (H0) is replaced by a
			computable halting floor. With (Ceff), one may also prescribe any
			fixed rational floor $\rho\in(0,1)$
			(Proposition~\ref{post:prop:ttrows}).
			
			\item A strict majority (C0) with positive halting probability (H0)
			gives exactly $\Delta^0_2$. The same class is obtained with any
			prescribed rational halting floor $\rho\in(0,1)$
			(Theorem~\ref{post:thm:main}).
		\end{enumerate}
		The constructions in all four parts are effective given the relevant
		programs, prescribed parameters, and witnesses for the computable
		guarantees. Proposition~\ref{post:prop:strict} gives the strict
		separations between the resulting classes.
	\end{theorem}
	
	\begin{proposition}\label{post:prop:strict}
		\[
		\text{decidable languages}
		\subsetneq \mathrm{btt}(\HALT)
		\subsetneq \mathrm{tt}(\HALT)
		\subsetneq \Delta^0_2.
		\]
	\end{proposition}
	
	\begin{proof}
		The one-query identity reduction for the undecidable language $\HALT$
		gives the first separation. For the middle one, write
		$x=\langle e,k\rangle$ using a computable pairing. At stage $t$, let
		$v(x,t)=0$ unless $P_e(x)$ has returned within $t$ steps a valid table
		with at most $k$ queries. If it has, evaluate that table with the answers
		observed by stage $t$. The appearance of the table and the discovery of
		positive answers cause at most $k+1$ changes in $1-v(x,t)$. Its limit
		therefore belongs to $\mathrm{tt}(\HALT)$ by Lemma~\ref{post:lem:bridge}.
		A proposed reduction of index $e$ and norm $n$ disagrees with this limit
		at $\langle e,n\rangle$. Proposition~\ref{post:thm:margin} gives the last
		separation.
	\end{proof}
	
	\subsection{Polynomially many coin tosses}\label{sec:fewcoins}
	
	A bound on random bits limits how often new halting runs can change
	the majority, even when running time remains unrestricted.
	
	\begin{proposition}[Polynomially many coin tosses]\label{post:rem:fewcoins}
		A language $L$ is decided under a strict conditional majority (C0) by a
		machine that tosses at most polynomially many coins on every run if
		and only if $L$ has a computable approximation that makes at most
		$2^{p(|x|)}$ mind changes on each input $x$, for some polynomial $p$.
		This class contains $\mathrm{btt}(\HALT)$ and is properly contained in $\mathrm{tt}(\HALT)$.
	\end{proposition}
	
	\begin{proof}
		We may take all polynomial bounds to have nonnegative integer
		coefficients, enlarging them if necessary. Suppose first that the
		machine uses at most $b=p(|x|)$ coin tosses on input $x$. Each of
		the $2^b$ strings of length $b$ determines a run, including when
		the run uses fewer than $b$ bits. At stage $t$, guess $1$ if
		$p_{1,t}(x)>p_{0,t}(x)$ and guess $0$ otherwise. Each change of
		guess requires at least one previously uncounted string to produce
		a halting output. Thus the guess changes at most $2^b$ times. The
		strict conditional majority ensures that it converges to
		$\one_L(x)$.
		
		Conversely, let $g$ be a computable approximation that makes at most
		$B=2^{p(|x|)}$ changes on input $x$. Toss $p(|x|)+2$ coins to choose
		uniformly among $4B$ branches. Assign one branch to the initial
		guess $g(x,0)$. For each $j=1,\ldots,B$, assign two branches to the
		$j$th change. Each of these branches computes the successive guesses
		until it sees that change, then halts with the new guess. All
		unassigned branches run forever.
		
		Initially the current guess leads by one branch. Whenever the guess
		changes, the new guess gains two branches and takes the lead by one.
		By induction, after the last change the correct bit leads by exactly
		one branch. All branches have the same probability, so this gives
		a strict conditional majority. The initial branch guarantees positive
		halting probability, and the machine uses no further coins after
		choosing its branch.
		
		A constant change bound satisfies an exponential bound of the stated
		form, so Lemma~\ref{post:lem:bridge}(1) gives the inclusion of
		$\mathrm{btt}(\HALT)$. The inclusion in $\mathrm{tt}(\HALT)$ follows from
		Lemma~\ref{post:lem:bridge}(2). To see that it is strict, enumerate
		pairs $(M_e,p_e)$ of machines and polynomial coin bounds. Cap $M_e$
		so that it runs forever if it attempts to toss more than
		$p_e(|x|)$ coins. On input $e$, identified with a binary string as
		above, compute the stage-by-stage guess for the capped machine and
		take its complement. This is a computable approximation with at most
		$2^{p_e(|e|)}$ changes on input $e$. The bound is computable from
		$e$, so the limit language belongs to $\mathrm{tt}(\HALT)$. If any
		polynomial-coin machine decided this language, it would appear with
		a valid bound at some index $e$. On that input its answer would be
		the complement of itself.
	\end{proof}
	
	\section{Quantum machines}\label{sec:quantum}
	
	For Deutschian computation with one level of closed timelike curves,
	Aaronson et al.\ obtained the same decision class, $\Delta^0_2$,
	for classical and quantum machines
	\citep[Theorems~3.2 and~3.3]{AaronsonEtAl2024}. For postselection
	on halting, we compare the probabilities of the individual outputs.
	We define a quantum machine with classical control and computable
	gates, and show that a classical probabilistic machine can match
	its two eventual output probabilities $p_0(x)$ and $p_1(x)$ exactly
	on every input. The converse simulation also preserves both
	probabilities. Their sum determines the halting probability, and
	their proportions determine conditional correctness. Thus the
	classifications above carry over to this quantum model with the
	same halting floors, confidence margins, and query bounds.
	
	\Needspace{8\baselineskip}
	\subsection{The model and its masses}
	
	\begin{definition}[Quantum machine]\label{post:def:quantum}
		A \emph{quantum machine} has a deterministic Turing-machine control
		with a read-only input tape, a classical work tape, and a register
		of qubits (quantum bits). The control computes qubit addresses on
		its work tape. When an instruction first names a qubit, the machine
		initializes it in the state $\lvert 0\rangle$. Thus the register is
		finite at every finite stage.
		
		Each step executes one instruction. An ordinary Turing-machine
		transition changes only the classical configuration. A gate
		instruction applies a gate from a fixed finite set $\mathcal G$ of
		unitary matrices to the named qubits. A measurement instruction
		measures one named qubit in the basis
		$\{\lvert 0\rangle,\lvert 1\rangle\}$ and branches on the outcome.
		A \emph{halting test} measures a named flag qubit in the same basis.
		On outcome $0$, the computation continues. On outcome $1$, the
		machine measures a named output qubit in that basis and halts with
		the resulting bit.
		
		The gate set $\mathcal G$ contains the Hadamard gate $H$ and the
		bit flip $X$. Its entries are uniformly computable: given a gate,
		a matrix position, and $k\in\N$, an algorithm returns rational
		approximations to the real and imaginary parts of that entry, each
		with error at most $2^{-k}$. A constant $c_{\mathcal G}$ bounds the
		number of qubits named by any one instruction.
	\end{definition}
	
	Fix an input $x$. A \emph{history} is a finite string $h$ of
	measurement outcomes. To follow the computation specified by $h$,
	use the next bit each time the machine measures a qubit. A halting
	test consumes one bit for the flag and, if the flag is $1$, one
	further bit for the output. A history is \emph{complete} if the
	machine halts exactly when it consumes the last bit of $h$.
	
	We compute the probability of a history by keeping its state vector
	unnormalized. Start with the all-zero register state of norm one.
	When a fresh qubit is named, append a factor $\lvert 0\rangle$ to
	the state. At a gate instruction, apply the gate to the named
	qubits and the identity to the others. At a measurement with
	recorded outcome $b$, apply the projector
	$\lvert b\rangle\langle b\rvert$ to the measured qubit and the
	identity to the others. If $\tilde u_h$ is the resulting vector,
	the probability of the recorded outcomes is
	$W(h)=\lVert\tilde u_h\rVert^2$.
	
	For $i\in\{0,1\}$, define
	\[
	p_{i,t}(x)=
	\sum_{\substack{h\text{ complete on }x\\
			h\text{ halts with output }i
			\text{ within }t\text{ steps}}}
	W(h).
	\]
	Distinct complete histories describe disjoint events, so this sum
	is the probability of halting with output $i$ by step $t$. As $t$
	increases, these probabilities converge from below to the eventual
	output probability $p_i(x)=\lim_{t\to\infty}p_{i,t}(x)$. The
	halting probability is $\Phalt(x)=p_0(x)+p_1(x)$. When $\Phalt(x)>0$,
	the conditional correctness for a language $L$ is
	\[
	q(x)=\frac{p_{\one_L(x)}(x)}{\Phalt(x)}.
	\]
	These quantities determine the answer rule and the guarantees in
	Definitions~\ref{ladder:def:halting} and~\ref{ladder:def:confidence}.
	
	\begin{lemma}[Masses]\label{post:lem:qmass}
		For every quantum machine, the finite-time output probabilities
		$p_{i,t}(x)$ are computable reals, uniformly in $x$, $t$, and $i$.
		For each $x$ and $i$, they are nondecreasing in $t$, and at every
		stage
		\[
		p_{0,t}(x)+p_{1,t}(x)\leq 1.
		\]
		Consequently, the eventual output probabilities $p_0(x)$ and $p_1(x)$
		are left-computably enumerable, uniformly in $x$.
	\end{lemma}
	
	The next lemma uses the fact that applying a Hadamard gate to a
	fresh qubit in state $\lvert 0\rangle$ and measuring it produces
	an independent fair bit.
	
	\begin{lemma}[Hadamard simulation]\label{post:lem:qcoins}
		Given the code of a classical probabilistic machine, one can compute
		the code of a quantum machine with exactly the same eventual output
		probabilities $p_0(x)$ and $p_1(x)$ on every input.
	\end{lemma}
	
	Appendix~\ref{app:quantum} proves both lemmas. The eventual output
	probabilities need not be computable uniformly in $x$. To see this,
	take the classical machine constructed in Theorem~\ref{post:thm:main}
	for an undecidable language in $\Delta^0_2$. If both output
	probabilities were computable uniformly in $x$, their sum $\Phalt(x)$
	would also be computable, and Proposition~\ref{post:prop:haltcomputable}
	would make the language decidable. Lemma~\ref{post:lem:qcoins} gives
	a quantum machine with the same output probabilities.
	
	\subsection{Classical simulation and transfer}
	
	\begin{lemma}[Classical simulation]\label{post:lem:qclassical}
		Given the code of a quantum machine, one can compute the code of a
		classical probabilistic machine with exactly the same eventual output
		probabilities $p_0(x)$ and $p_1(x)$ on every input.
	\end{lemma}
	
	\begin{proof}
		To apply Lemma~\ref{post:lem:realize}, we need nondecreasing rational
		approximations to the two output probabilities. Fix $x$ and
		$i\in\{0,1\}$. At stage $s$, use Lemma~\ref{post:lem:qmass} to
		compute a rational $r_{i,s}$ within $\varepsilon_s=2^{-s-2}$ of
		$p_{i,s}(x)$. Then $c_{i,s}=r_{i,s}-\varepsilon_s$ is a lower
		bound satisfying
		\[
		p_{i,s}(x)-2\varepsilon_s \leq c_{i,s}\leq p_{i,s}(x).
		\]
		Keep the largest lower bound obtained so far:
		\[
		\ell_i(x,t)=\max\{0,c_{i,0},\ldots,c_{i,t}\}.
		\]
		These values are computable rationals and are nondecreasing in $t$.
		Since $p_{i,s}(x)\leq p_{i,t}(x)$ for $s\leq t$, we have
		$\ell_i(x,t)\leq p_{i,t}(x)$. In particular,
		\[
		\ell_0(x,t)+\ell_1(x,t)
		\leq p_{0,t}(x)+p_{1,t}(x)\leq 1.
		\]
		Thus $(\ell_0,\ell_1)$ is a mass program. Its limits are the required
		output probabilities, since
		\[
		p_{i,t}(x)-2\varepsilon_t
		\leq \ell_i(x,t)\leq p_{i,t}(x)
		\]
		and both bounds converge to $p_i(x)$. Lemma~\ref{post:lem:realize}
		now gives a classical machine with exactly these eventual output
		probabilities.
	\end{proof}
	
	\begin{theorem}[The classes are unchanged]\label{post:thm:quantum}
		The classifications of Sections~\ref{sec:postselection}
		and~\ref{sec:ladder} by halting and confidence guarantees hold for
		quantum machines with the same
		halting floors, confidence margins, and effective constructions.
		In particular, the sharp query bound in
		Theorem~\ref{post:thm:sharp} is again $n(\rho,\eta)$.
	\end{theorem}
	
	\begin{proof}
		By Lemmas~\ref{post:lem:qcoins} and~\ref{post:lem:qclassical}, the
		conversion in either direction preserves $p_0(x)$ and $p_1(x)$ on
		every input. It therefore
		preserves their sum $\Phalt(x)$ and, whenever this sum is positive,
		the conditional correctness $q(x)$. The same floor or margin witnesses
		the corresponding guarantee for the converted machine. In case (Hc),
		the same program still approximates $\Phalt(x)$ to the requested
		precision. Both conversions are effective.
	\end{proof}
	
	The classical simulation lets us reuse the upper-bound proofs even
	when the quantum gates have arbitrary computable entries. In that
	setting the finite-time output probabilities are computable reals,
	and an exact comparison of two such probabilities need not be
	decidable. Lemma~\ref{post:lem:qclassical} uses rational lower bounds
	to recover the eventual output probabilities. For the lower bounds,
	applying a Hadamard gate to a fresh qubit and measuring it supplies
	each fair bit the classical construction needs.
	
	\section{Comparison with Deutsch's rule and open questions}
	\label{sec:discussion}
	Table~\ref{post:tab:rules} compares the two rules. In the polynomial-time
	rows, postselection conditions on a designated measurement outcome. In
	the unrestricted rows it conditions on eventual halting.
	
	\begin{table}[tbp]
		\centering\footnotesize
		\caption{Deutschian computation and postselection across resource
			and error bounds. PSPACE denotes polynomial space, and PP denotes
			unbounded-error probabilistic polynomial time.}
		\label{post:tab:rules}
		\begin{tabularx}{\linewidth}{@{}
				>{\raggedright\arraybackslash}p{.13\linewidth}
				>{\raggedright\arraybackslash}p{.15\linewidth}
				>{\raggedright\arraybackslash}p{.12\linewidth}
				>{\raggedright\arraybackslash}X
				>{\raggedright\arraybackslash}X@{}}
			\toprule
			Time bound & Loop width & Error & Deutsch's rule & Postselection \\
			\midrule
			Polynomial & Polynomially many bits or qubits & Bounded &
			Classical and quantum: PSPACE \citep{AaronsonWatrous2009}. &
			Quantum: PP \citep{Aaronson2005}. See also
			\citep[Section~9]{AaronsonEtAl2024}. \\
			\addlinespace
			Polynomial & One bit or qubit, up to logarithmic width & Bounded &
			One classical loop bit gives the postselected class of the
			underlying machine \citep{SayYakaryilmaz2011UC,SayYakaryilmaz2012NatComp}.
			Logarithmically many loop bits give the same power
			\citep{ODonnellSay2014}. With logarithmically many loop qubits, the
			quantum class is PP \citep{ODonnellSay2018}. &
			One loop qubit suffices to implement a postselected measurement
			\citep{BrunWilde2012}. \\
			\midrule
			Unrestricted & Unbounded & Zero &
			$\Delta^0_2$ \citep[Theorems~3.2 and~3.3]{AaronsonEtAl2024}. &
			Decidable (Proposition~\ref{post:prop:zeroerror}). \\
			\addlinespace
			Unrestricted & Unbounded & Bounded &
			$\Delta^0_2$ \citep[Theorems~3.2 and~3.3]{AaronsonEtAl2024}. &
			Fixed margin: $\mathrm{tt}(\HALT)$
			\citep[Proposition~9.2]{AaronsonEtAl2024}
			(Proposition~\ref{post:prop:ttrows}).
			Prescribed floor and margin: at most $n(\rho,\eta)$ nonadaptive
			queries (Theorem~\ref{post:thm:sharp}). \\
			\addlinespace
			Unrestricted & Unbounded & Unbounded &
			$\Delta^0_2$ \citep[Theorems~3.2 and~3.3]{AaronsonEtAl2024}. &
			$\Delta^0_2$, also with any prescribed rational floor $\rho<1$
			(Theorem~\ref{post:thm:main}). \\
			\bottomrule
		\end{tabularx}
		
		\smallskip
		\begin{minipage}{\linewidth}
			\footnotesize
			For postselection on halting, almost-sure halting gives decidability
			even under a strict majority
			(Proposition~\ref{post:prop:haltcomputable}). The almost-surely
			halting model with an explicit discard output also gives decidability
			\citep[Proposition~9.1]{AaronsonEtAl2024}. All unrestricted
			postselection entries apply to both classical machines and the
			quantum machines of Section~\ref{sec:quantum}, with the same
			constants (Theorem~\ref{post:thm:quantum}).
		\end{minipage}
	\end{table}
	
	With unrestricted time and loop width, a fixed margin already separates
	the two rules. Deutsch's rule still gives $\Delta^0_2$, while
	postselection on halting, with positive halting probability on every
	input and a fixed margin $0<\eta<1/2$, gives $\mathrm{tt}(\HALT)$, a
	proper subclass
	(Propositions~\ref{post:thm:margin} and~\ref{post:prop:ttrows}).

	A machine may also ask membership questions about the language decided
	by another machine. If every question receives the exact membership
	bit, the outer machine is a probabilistic machine with an oracle, and
	the arguments of Sections~\ref{sec:postselection} and~\ref{sec:ladder}
	relativize. The first version of our preprint carries this out for
	machines nested to any finite depth~\citep[Sections~5.2 and~5.3]{FarhatDidehvar2026v1}.
	At depth $k$, a strict majority then gives the languages decidable
	with an oracle for the halting problem iterated $k$ times. Raat
	obtained the same class for nested Deutschian loops already with
	bounded error~\citep[Section~5.3]{Raat2023}\,\citep[Remark~3.4]{AaronsonEtAl2024}.
	
	Consider a classical polynomial-time machine that outputs $0$, outputs
	$1$, or discards its run. Require the correct bit to have a strict
	majority among the runs it keeps. Replacing each discard by a fair bit
	preserves the difference between the two output probabilities. Conversely,
	a $\mathbf{PP}$ machine whose acceptance probability never equals $1/2$
	can keep every run. Every language in $\mathbf{PP}$ has such a
	machine: if a machine tosses $m$ coins, run it with probability $1/2$
	and otherwise accept with probability $1/2-2^{-m-1}$. The new
	acceptance probability exceeds $1/2$ exactly when the old one did,
	and it never equals $1/2$. This gives exactly
	$\mathbf{PP}$, also relative to an oracle.
	
	Write $\mathbf{NP}$ for nondeterministic polynomial time and
	$\mathbf{P}^{\mathbf{NP}}$ for deterministic polynomial time with
	adaptive queries to an $\mathbf{NP}$ oracle.
	Beigel~\citep{Beigel1994} constructed an oracle relative to which
	$\mathbf{P}^{\mathbf{NP}}\not\subseteq\mathbf{PP}$. In that setting,
	some languages decidable with adaptive $\mathbf{NP}$ queries lie outside
	strict-majority postselection. In the unrestricted setting,
	Theorem~\ref{intro:thm:A} shows that postselection on halting captures
	every adaptive use of $\HALT$.
	
	\subsection{Open questions}
	
	\paragraph{Sampled oracles.}
	Let $N$ be a machine as in Definition~\ref{post:def:model}, and let
	$M$ be a probabilistic machine that may ask questions $y$ about $N$.
	Suppose that, instead of the exact membership bit, each question
	returns a fresh sample from the outputs of $N$ on $y$ that halt. Write $p_b^N(y)$ for the
	probability that $N$ halts on $y$ with output $b$, and
	$\Phalt^N(y)=p_0^N(y)+p_1^N(y)$. Given the current query $y$ and
	the computation so far, the reply equals $b$ with probability
	\[
	\frac{p_b^N(y)}{\Phalt^N(y)}.
	\]
	By definition, each oracle call returns an independent draw from this
	conditional output distribution of $N$ on $y$. Every guarantee asked
	about below requires $\Phalt^N(y)>0$, which makes the distribution
	well defined, although an ordinary run of $N$ may still diverge. The
	reply uses fresh randomness, including when $M$ repeats a query. The output probabilities of $M$ are
	taken over both its own coins and the sampled replies.
	
	Which languages can the pair $(M,N)$ decide when both machines must
	give a strict conditional majority? How does the class change when
	the nesting is deeper, or under the floor and confidence guarantees
	of Section~\ref{sec:ladder}?

	\FloatBarrier
	\begingroup\small
	\bibliographystyle{unsrtnat}
	\bibliography{references}
	\endgroup
	
	\Needspace{12\baselineskip}
	\appendix
	\section{Proofs for Section~\ref{sec:quantum}}\label{app:quantum}
	
	\begin{proof}[Proof of Lemma~\ref{post:lem:qmass}]
		Fix an input $x$. Follow the computation for $t$ steps, keeping each
		halted branch as a leaf and extending each active branch according
		to its next instruction. An ordinary transition preserves the weight
		of a branch, as does a unitary gate. A measurement splits the branch
		into two children whose weights sum to the parent's weight: for an
		unnormalized state $u$ and the two measurement projectors $E_0,E_1$,
		\[
		\lVert E_0u\rVert^2+\lVert E_1u\rVert^2=\lVert u\rVert^2.
		\]
		A halting test first splits the branch according to the flag and
		then splits the flag-$1$ branch according to the output bit. Thus
		the total weight of the leaves remains one. Extending the tree
		preserves every halted leaf, so $p_{i,t}(x)$ is nondecreasing in
		$t$, and $p_{0,t}(x)+p_{1,t}(x)\leq 1$.
		
		To compute $p_{i,t}(x)$, construct this finite tree by following
		both outcomes of every measurement, including outcomes of weight
		zero. The control is deterministic once the outcomes are specified,
		so the tree can be constructed without testing whether a branch
		has positive weight. Along each branch there are at most $2t$
		measurement outcomes, counting the flag and output bits of each
		halting test, and at most $c_{\mathcal G}t$ named qubits.
		
		For each leaf that halts with output $i$, compute its weight as a
		finite expression in the gate entries. Starting from the all-zero
		state, apply the gates and projectors along the branch, then sum
		the squared moduli of the resulting amplitudes. Adding these
		expressions over the selected leaves gives $p_{i,t}(x)$. The
		expression uses only addition, multiplication, and complex
		conjugation. To approximate its value within $2^{-k}$, enclose the
		real and imaginary parts of the gate entries in rational intervals
		and evaluate the expression by interval arithmetic. Refine the
		entry intervals until the resulting interval for $p_{i,t}(x)$ has
		width at most $2^{-k}$. This search terminates because the
		expression is finite and its operations are continuous. The
		procedure is uniform in $x,t,i,k$.
		
		Finally, choose rationals $r_{i,s}$ with
		$|r_{i,s}-p_{i,s}(x)|\leq 2^{-s}$. The running lower bounds
		\[
		\ell_i(x,t)
		=\max\bigl(\{0\}\cup\{r_{i,s}-2^{-s}:0\leq s\leq t\}\bigr)
		\]
		are computable and nondecreasing. They satisfy
		\[
		p_{i,t}(x)-2^{1-t}\leq \ell_i(x,t)\leq p_{i,t}(x).
		\]
		Since $p_{i,t}(x)\to p_i(x)$, these rational lower bounds have the
		same limit. Thus $p_i(x)$ is left-computably enumerable uniformly
		in $x$.
	\end{proof}
	
	\begin{proof}[Proof of Lemma~\ref{post:lem:qcoins}]
		Copy the classical control. To simulate a
		coin toss, apply $H$ to a fresh qubit initialized in
		$\lvert 0\rangle$, measure it in the computational basis, and take
		the classical transition specified by the outcome. Since
		\[
		H\lvert 0\rangle
		=\frac{\lvert 0\rangle+\lvert 1\rangle}{\sqrt{2}},
		\]
		each outcome has probability $1/2$ conditional on every preceding
		history. Using a fresh qubit for each toss therefore gives the
		independent fair bits required by the classical machine.
		
		When the classical machine halts with bit $i$, prepare a fresh output
		qubit in state $\lvert i\rangle$ by applying $X$ to it if $i=1$.
		Apply $X$ to a fresh flag qubit and execute the halting test. The
		flag is then $1$ and the measured output is $i$, both with
		probability one.
		
		A classical run that halts after consuming a particular sequence
		of $\ell$ coin bits has probability $2^{-\ell}$. Its corresponding
		quantum history has the same probability and output. Conversely,
		every quantum history that halts with positive weight arises from
		such a classical run. Summing over the halting runs gives the same
		eventual masses $p_0(x)$ and $p_1(x)$ on every input. Replacing
		the instructions as described computes the quantum machine's code
		from the classical machine's code.
	\end{proof}
\end{document}